# In-Storage Embedded Accelerator for Sparse Pattern Processing

Sang-Woo Jun[*], Huy T. Nguyen[#], Vijay Gadepally[#*], and Arvind[*]
[#]MIT Lincoln Laboratory, [*]MIT Computer Science & Artificial Intelligence Laboratory
Email: wjun@csail.mit.edu, hnguyen@ll.mit.edu, vijayg@ll.mit.edu, arvind@csail.mit.edu

*Abstract* We present a novel architecture for sparse pattern processing, using flash storage with embedded accelerators. Sparse pattern processing on large data sets is the essence of applications such as document search, natural language processing, bioinformatics, subgraph matching, machine learning, and graph processing. One slice of our prototype accelerator is capable of handling up to 1TB of data, and experiments show that it can outperform *C/C++* software solutions on a 16-core system at a fraction of the power and cost; an optimized version of the accelerator can match the performance of a 48-core server.

## I. Introduction

Many data analysis algorithms of interest on large data sets composed of documents, images, audio and video can be formulated as operations on very large but very sparse vectors and matrices. There are two challenges: the size of data is generally too large to fit in the system memory (DRAM) of a single server [1], and current server architectures are far from ideal for processing sparse datasets, causing the CPU itself to become a bottleneck.

The traditional way to overcome the size challenge is to use a cluster of machines so that data can be accommodated in the collective main memory (DRAM), and distribute the computation across the machines in the cluster [2]. A 1.5 TB 192-core distributed system with a dozen nodes of 128 GB DRAM memory each would cost about $60k. This system would use a software layer for distributed data processing such as Hadoop. With circa 2016 new server technology, it is possible to configure memory capacity up to 1.5 TB on a quad-socket server, which would cost $27k to $49k, for 48 cores and 72 cores, respectively. This single-box system would not need to access data over the network, thus, its software could be streamlined for performance. The industry's trend of adding DRAM and cores per server helps consolidating the machines and improves performance. However, it is not a panacea. Large memory buffers cause large loading, and requires high power circuitries for fast access. Data and results also need to be stored on non-volatile storage devices for disruption recovery.

Flash-based secondary storage such as Solid-State Drives (SSDs) is a high performance and power efficient technology solution. A recent interface standard, the Non-Volatile Memory Express (NVMe), makes this type of solution even more attractive. We propose to take this technology further: to use flash-based non-volatile memory (NVM) as main data store instead of DRAM. Comparatively, flash is at least 10x cheaper, takes 10x less space, and is 10x less power-consuming than DRAM. In such a system, all data will be on SSDs for processing rather than read in from hard drives. Of course, using flash memory in place of DRAM incurs longer latency in accessing information, and consequently the system has to be optimized for such accesses. Instead of approaching this as a storage problem, we will address it in conjunction with computing, and optimize across boundaries where possible to achieve a better overall solution.

Computing with application-specific hardware accelerators can lead to one to three orders of magnitude better performance with less power consumption compared to CPU cores performing similar tasks [3]. Many accelerator devices are packaged as an independent system component, which can be plugged into a high-speed bus such as PCIe to interface with CPUs and system memory.

In this paper, we demonstrate a system that integrates the storage and computing solutions together, as shown in Fig. 1, to reduce DRAM memory size and CPU workload [4]. In this architecture, the Field-Programmable Gate Array (FPGA) directly accesses flash storage, and processes data prior to presenting the results to the CPU, hence, "in-storage computing". A dedicated high bandwidth/low-latency FPGA-based network is used to connect FPGAs together for scaling to problem size of 10s of TBs.

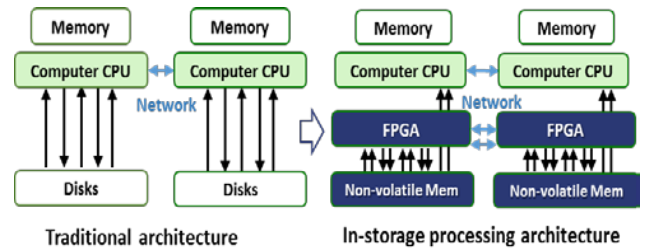
Fig. 1. In-Storage Computing Architecture

Sparse pattern processing is very inefficient on general purpose computers, and is a good candidate for acceleration on our in-storage computing architecture. We will show that our baseline accelerator outperforms a 16-core server system while using only 2/3 power, and an optimized version could match a 48-core system at ¼ power and ¼ cost. (These comparisons are based on C/C++ server code; comparisons with Java-based implementation would correspond to 3x more cores, i.e., 48 cores and 144 cores.) Each accelerator slice can handle problem size of 1 TB.

This work is partially supported by the Assistant Secretary of Defense for Research and Engineering under Air Force Contract number FA8721-05-C-0002. Opinions, interpretations, conclusions and recommendations are those of the author and are not necessarily endorsed by the United States Government.

We will discuss how sparse pattern processing operations can be applied to many applications such as document search, natural language processing, bioinformatics, subgraph matching, machine learning, and graph processing. [5][6][7]. A recent effort, the GraphBLAS [8], aims to standardize some of these computational kernels in order to support the development of hardware-accelerators [9].

*Paper organization:* Section II describes sparse pattern matching using document search as an example and presents several important related applications. Section III introduces our architecture. Section IV describes implementation details of the prototype system and its performance measurements. Sections V and VI conclude with discussions and summary.

## II. SPARSE PATTERN MATCHING

We begin by showing how document search problem can be formulated as sparse vector multiplications, and then discuss applications that can benefit from the sparse processing kernels developed for document processing.

### A. Document Matching

The objective of document matching is to find document candidate(s) that match best to a query document [10][11][12]. The key idea is similar topics would use similar vocabulary at high level of occurrences and can be abstracted into mathematical models based on the words, their frequencies, and with more sophistication, the ordering and grouping of the word appearances. Topic modeling occurs in applications such as natural language understanding, relationship extraction, sentiment analysis, topic segmentation, information retrieval, predictive analysis, and bioinformatics.

A simple example of a document search is shown in Fig. 2, where documents A and B have been pre-processed to extract prominent words with occurrences above some threshold. In the UCI Machine Learning Repository [13], the collection of NY Times articles contains around 300,000 documents with about 100,000 prominent words. The Enron email collection has almost 40,000 emails with about 28,000 prominent words.

Fig. 2. Document Matching

**Comparison metric - Cosine similarity:** Each document can be represented as an *N*-dimensional vector, where *N* is the size of the bag-of-words. Each word in the vocabulary is assigned to one of the *N* dimensions, and the value of a dimension corresponds to the occurrence frequency of the assigned word. In the above example, the document-A vector would be stored as a sparse vector of 4 non-zero elements, A = [$A_1$, $A_3$, $A_7$, $A_9$]. The indices 1, 3, 7, 9, correspond to the word indices in the bag-of-words. Document-B vector also happens to have 4 non-zero elements, B = [$B_1$, $B_2$, $B_7$, $B_{10}$].

One simple and effective method for comparing high-dimensional vectors is the *Cosine similarity metric*, which is defined below (the top computes the correlation, and the bottom performs the normalization). Vectors that are closely aligned would result in large Cosine metric.

$$\cos(\theta) = \frac{\sum_{i=1}^{N} A_i B_i}{\sqrt{\sum_{i=1}^{N} A_i^2} \sqrt{\sum_{i=1}^{N} B_i^2}}$$

Fig. 3 illustrates the computation for the correlation of sparse vectors A and B, where the vectors A and B correspond to documents A and B presented above. The element-wise multiplication of $A_i$ and $B_i$ creates partial products $PP_i$ for each term, but due to sparsity, only the terms $PP_1$ and $PP_7$ are nonzero and need to be created. The correlation score is the summation of these partial products.

Fig. 3. Pattern Matching as Sparse Vector Multiplication

The match processing of a document-A against the whole dataset, i.e., document-B, document-C, etc., can be formulated as a matrix-vector multiplication as illustrated in Fig. 4. The sparse matrix is the collection of all sparse vector representations of the documents in the dataset.

Fig. 4. Document Search and Classification

There are many opportunities for parallelizing the processing. One could partition the matrix into *K* subsets (separated by dividing lines in Fig. 4) and compute in parallel on *K* accelerator kernels.

If a batch of *L* queries is issued, the problem can be formulated as sparse matrix - sparse matrix multiplication. The work can be performed in parallel on *K\*L* kernels, with *K* and *L* parameters determined for optimal accelerator performance, based on the accelerator computing capability, data bandwidth, and local working storage.

## B. Other applications

**Subgraph Matching:** Document matching could be extended into the graph processing domain by considering subgraphs to be documents, and conducting a search for subgraphs that contain similar edges. As illustrated in Fig. 5, each edge could be represented as a "word" containing the labels of its two vertices. Subgraphs can have different number of edges, just like documents can have different sizes. Traversing an edge in the graph corresponds to performing a word matching operation. The labeling and partitioning of edges into subgraphs depends on the application context. Our focus here is on the acceleration of subgraph matching [14], which is important in biological networks, event recognition, and community detection.

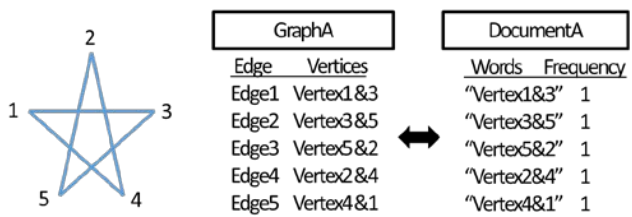

Fig. 5. Subgraph Matching

**Feature Matching in Machine Learning:** Machine learning algorithms train their classifiers by churning through massive amount of data to evaluate and optimize a set of coefficients, so that the classifications aligns to desired outcomes. A facial recognition application, for example, would operate on feature descriptors extracted from facial images, or prominent words in a document search analogy. The training process involves many rounds of matching faces against classifier candidates, not too different from the scenario depicted in Fig. 4.

For applications that use neural networks, sparse representation and evaluation enables efficient implementation. A low-power, portable form factor accelerator could allow the training of neural networks to take place in the field rather than back at the office.

**Bioinformatics**: High dimensional search using a bag-of-words representation is also useful in the field of bioinformatics. For example, researchers use protein search tools such as BLAST[15] to guess the genealogy and function of proteins by matching against previously annotated proteins. Since performing an optimal search algorithm such as Smith-Waterman on the entire annotated protein database is expensive, each protein sequence in the database is pre-processed into a more easily searchable format, such as bag-of-words. This pre-processing step is not to find the most similar sequence, but rather, to determine a smaller set of reference sequences statistically likely to be similar to the query, so that the search space of the optimal algorithm is significantly reduced.

We have experimented with this approach, by first pre-process each protein sequence to generate a bag-of-words representation of all 3-mers, as shown in Fig. 6. Once all reference protein sequences in the dataset are processed and encoded in such a sparse format, the previously described document search kernel can be used to quickly find the set of relevant reference proteins. This reduces the search on the 35 GB UniProt TrEMBL dataset [16] to an approximately 4 GB bag-of-words, which requires only 2 seconds for our flash-based prototype to traverse.

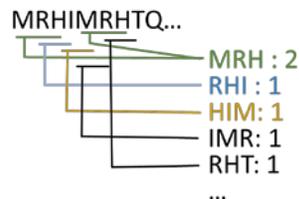

Fig. 6. Organizing a Protein Sequence into Bag-of-Words

## III. ACCELERATOR DESIGN

As part of the design process, we performed analysis and experiments to gain insights into the application. By decomposing the application into key functions and performing benchmarking, we can tell if performance is bounded by data bandwidth or computing capability.

For document matching, computing the Cosine similarity metric is the key function. Specialized accelerator can significant improve the performance. Fast data access is required to keep its computing kernels busy. It is, therefore, critical to use data structures that are as bandwidth-efficient.

### A. Data Representation

Rather than storing the words, we store the index into the bag-of-words (see Fig. 7). This requires an indexing step, but allows for much flexibility in abstraction, for example, "words" can be phrases, strings, or even other features (images) of the document, etc.

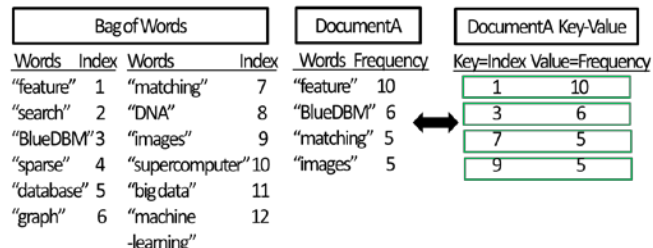

Fig. 7. Document Encoding Schema

To make effective use of storage bandwidth and computation, we encoded the vectors, or pattern datasets, into binary format as shown in Fig. 8. Each data item is 32 bits wide, and is either a pattern identifier, i.e., "Document A," or a key/value pair for each word in the document. Each document in a large dataset can be encoded with a single pattern identifier followed by a list of key/value pairs. Comparing to the original data format from UCI repository [13], which replicates the documentID with each wordID and stores one word per line, this format saves almost 50% storage bandwidth.

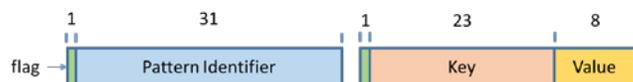

Fig. 8. Data Format

## B. Architecture

An architectural view of the sparse pattern matching kernel is shown in Fig. 9. It consists of a query memory, accessed through the sideband datapath, and a chain of computation modules for calculating the Cosine similarity between the query and data from flash storage. Multiple kernels are implemented in the system in order to make full use of the flash storage bandwidth. Each kernel receives its own data routed from flash storage as controlled by the host software. Data pages from flash do not necessarily arrive in order, which necessitates re-ordering and low-level coordination activities that takes place in the flash storage interface logic block [4].

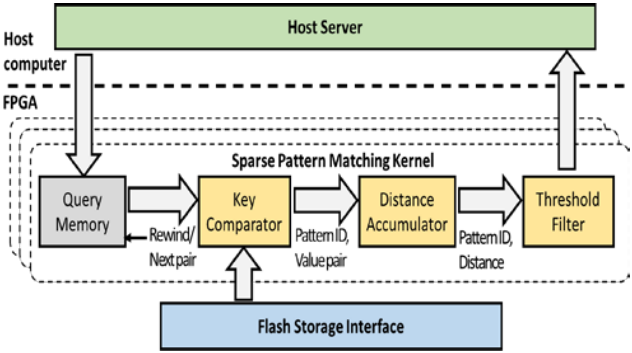

Fig. 9. Sparse Matching Accelerator Architecture

The Key Comparator module performs matching of words from the query document against words of all documents in the target dataset. A match is declared when the wordID in a document is the same as the wordID from the query. This step identifies the index $i$ of the nonzero partial product $PP_i$ in the sparse multiplication described in Fig. 3. The generation of $PP_i$ and computation of Cosine similarity score are performed in the subsequent Distance Accumulator module. The score is accumulated for each document. At the end, documentIDs with high scores are reported to the computer. Throughout the processing, subsets of wordIDs are matched in parallel by many kernels. Due to sparsity, many comparisons must be performed until a successful match. Thus, the performance of the Key Comparator has significant impact to the overall operation.

The Query Memory uses a prefetch predictor and advanced logic to keep the processing pipeline full. A pointer is used to access the query key (in query memory). A second pointer tracks the (sorted) candidate key after loading from flash. For each comparison, one of the two pointers is incremented, depending on which is bigger. The candidate key pointer only increases, but the query memory pointer occasionally needs to be "rewound" upon the end of the candidate document is reached. Without the prefetch predictor, the loading of a new query key from query memory would occur only after it can be determined that no rewinding is required. Since loading involves latency, this would waste valuable cycles and reduce performance. Adding the prefetch predictor logic, as shown in Fig. 10, allows the query memory to queue prefetched keys into the comparison unit.

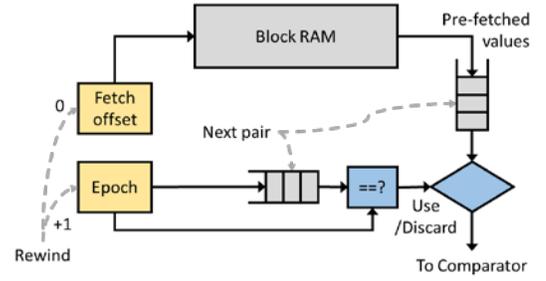

Fig. 10. Prefetching of Query

The prefetcher assumes rewind is not to happen. It keeps requesting reads from block RAM, and enqueues an epoch value for each read request. The epoch value provides a frame of reference to distinguish one document processing interval from another. When a rewind happens, the prefetcher increments the epoch value. Previously enqueued prefetches are considered mispredicted reads and discarded. They are identified by comparing the enqueued epoch values with the current epoch value.

Fig. 11 shows the interaction of host software accelerator and flash storage. Each FPGA accelerator kernel implements the same interface and can be accessed through 4 ports: dataIn, commandIn, resultToMemory, and dataToStorage. Data read from flash storage is streamed through dataIn. Commands from the host are sent via the sideband port commandIn. As the kernel performs computation, it can send data to be stored to flash via dataToStorage, or to host software via resultsToMemory. The host software is responsible for managing the ports, for example, maintaining a list of free pages that dataToStorage uses.

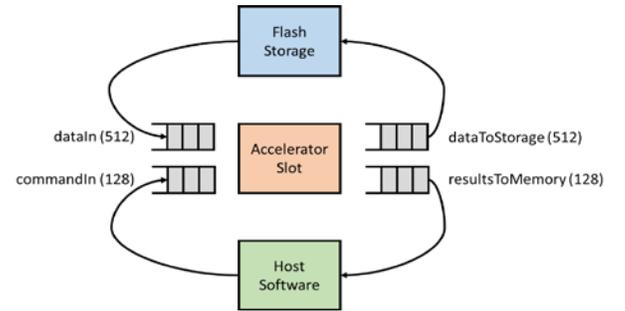

Fig. 11. Accelerator Interface

**Software:** A reference C/C++ software implementation was developed to provide benchmark functionality and a framework for later integration of the accelerator. The implementation is parameterized so the user can choose the number of worker threads to spawn. Each worker is given its own copy of the query pattern and a contiguous partition of the dataset. After finishing execution, each thread reports its top pattern with the largest Cosine metric. The main thread then selects the global nearest pattern from the small list of local patterns.

## IV. IMPLEMENTATION & RESULTS

### A. System description

Our application was implemented on a single-node setup of MIT's BlueDBM system [4]. A node consists of a 24-core, 50 GB DRAM Xeon server and a BlueDBM storage device, which includes a Xilinx VC707 FPGA development board and a custom-built 1 TB flash module capable of 2 GB/sec throughput. The BlueDBM device and host server are connected via a Gen2 x8 PCIe link. Fig. 12 highlights one "accelerator slice" on a server computer and the software infrastructure for integrating the accelerator.

Eight accelerator kernels were used to fully take advantage of the 2 GB/s flash bandwidth. Each kernel was given 8 KBs of query memory on the FPGA block RAM to store a sparse query vector of up to 2K nonzero elements. The size of the query memory could be made larger and more accelerator kernels could be used, but this was more than adequate for our example application.

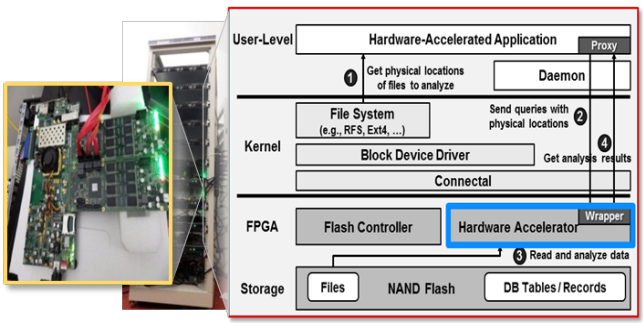

Fig. 12. BlueDBM System

We used UCI's bag-of-words dataset [13] and created bag-of-words databases for multiple document collections, including Pubmed. Each database consists of a vocabulary file that maps words to wordIDs, and a doc file that contains {documentID, wordID, wordCount} tuples. We also created large datasets of 100 GBs or more for experiments by developing a synthesizer that generates permutations of the smaller sets through adding and removing random words, and assigning random count values to some words.

### B. Results

The document matching performance was averaged over an interval of 10 seconds for four system configurations:

(1) Hard Disk: Data on hard disk
(2) RAM-Disk: Data in RAM - avoids the mechanical operation of the disk but still requires the operating system to perform file operations. Approximates the upper-bound performance of SSDs.
(3) Memory: Represents the new trend of in-memory database processing enabled by the availability of large computer memories and CPU cores.
(4) BlueDBM: Data in flash and most operation off-loaded to the FPGA accelerator. Minimizes load on CPU and memory.

Fig. 13 shows the performance for various configurations as the number of CPU threads is scaled from 1 to 24. Configuration (1) with hard disk is clearly limited by the disk bandwidth. Configuration (2) using RAM disk scales to 6 million docs/sec, but eventually runs into the limitation of file processing by the OS. Configuration (3) with all data in memory achieves the highest rate of 13 million docs/sec for 24 threads. In real-life, the computer would need to read data from secondary storage into memory at least once, so performance would be somewhere in between configurations (2) and (3).

Configuration (4) using BlueDBM reaches its peak at **10 million docs/sec**, limited by the bandwidth of the BlueDBM flash modules. It is slightly lower than in-memory processing but almost twice the performance of RAM disk. Since this configuration offloads most of processing onto the FPGA, the host software consists of only two threads for managing the file system and the computation. The CPU threads >2 in Fig. 13 only apply to non-BlueDBM systems, as BlueDBM requires two threads and no more.

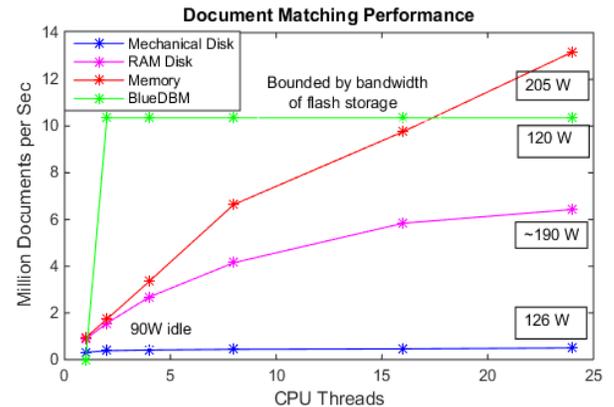

Fig. 13. Document Processed vs CPU Threads

Note that BlueDBM efficiency comes from our in-store processing paradigm that uses the FPGA for both computing acceleration and storage control. Had the FPGA been used for acceleration alone, the host server would still have required many threads for 2 GB/s data transfer between storage and accelerator. With BlueDBM achieving 10 million docs/sec using only 2 CPU threads, a possibility opens up for extending the application into low-power portable use cases. A much smaller CPU, such as those embedded in the FPGA could be used, which prompts an interesting concept of shrinking and folding the host into the accelerator itself!

The power measurements are given in Table 1, ranging from 90 Watts idle to 205 Watts by configuration (3) for in-memory processing. Configuration (4) with BlueDBM draws only 120 W, nearly 40% less power, while delivering a slight reduction in processing rate compared to in-memory processing. It is also 40% more power efficient and has 40% better performance than configuration (2) using RAM disk (upper-bound for SSDs), as shown in Fig. 13.

Table 1: Power Dissipation

| Number of CPU threads | CPU – C++ | | (4) BlueDBM with 8 comparators |
|---|---|---|---|
| | (1) Data in hard drive | (3) Data in memory | |
| 1 | 120 | 114 | NA |
| 2 | 110 | 134 | 120 W (idle 90W) |
| 4 | 100 | 145 | |
| 8 | 123 | 160 | |
| 16 | 112 | 164 | |
| 24 | 126 | 205 | |

## V. DISCUSSION

### A. Scalability & Cost

The performance of BlueDBM accelerator is limited by the bandwidth between the accelerator kernels and the flash storage, which based on the measured figure of 10 million docs/sec is about 2 GB/s. For the same storage bandwidth, it is possible to increase computation by processing several queries together in a batch. Recall the sparse matrix – sparse matrix discussion near Fig. 4. Up to 20 kernels could be implemented in the FPGA to process **3 queries in parallel**, yielding an estimated rate of **27 million docs/sec** as shown in Table 2. This performance could alternatively be achieved for single query issue if the storage bandwidth is increased to 6 GB/sec, by either accessing neighboring BlueDBM flash modules via the FPGA network (see Fig**.** 1), or upgrading the architecture to use multiple PCIe-based modules.

Table 2: Accelerator Scalability

| Kernels in FPGA | Million Docs / sec | Flash IO (GB/s) |
|---|---|---|
| 8 | 10.35 | ~2 |
| 20 | 27 est. | 5.4 est. |

The CPU in-memory processing, in contrast, is not bandwidth but rather processing limited. The only way to get more processing is adding more cores. Extending Fig. 13 to 27 million docs/sec reveals that a **48-core server** system is needed. With 2016 technology, a quad-socket platform could meet this requirement with 12-core CPUs.

Such a system with ~1.5 TB of memory costs about $27k, with approximately $10k for memory, $7k for platform and storage, and $10k for the four 12-core CPUs (or $32k for four 18-core CPUs). An FPGA-based accelerator slice including flash modules can be built from commercial components for less than $6k, or **¼ the cost**.

### B. Power Optimization

Fig. 14 captures performance and power dissipation for configurations (3) and (4), where the solid red curve depicts in-memory CPU processing, and the solid green curve for BlueDBM in-storage processing. The dotted curves project the capability of a slightly modified BlueDBM system as discussed in Table 2.

The blue dotted curve shows an interesting variant of BlueDBM where the host CPU is downsized and absorbed into the accelerator itself, i.e., the host-side 2-thread software now runs on an embedded CPU within the FPGA. This presents a solution for extreme Size, Weight, and Power (SWaP) applications.

The idle SWaP solution would draw only 10 Watts rather than 90 Watts. The blue curve is obtained by shifting the green curve down by 80 Watts. At the 10 million doc/sec, power dissipation is expected to be < 50 Watts, and at 27 million docs/sec about 90 Watts. Our embedded in-storage processing accelerator would be about **3.8 times more power efficient** than in-memory CPU processing.

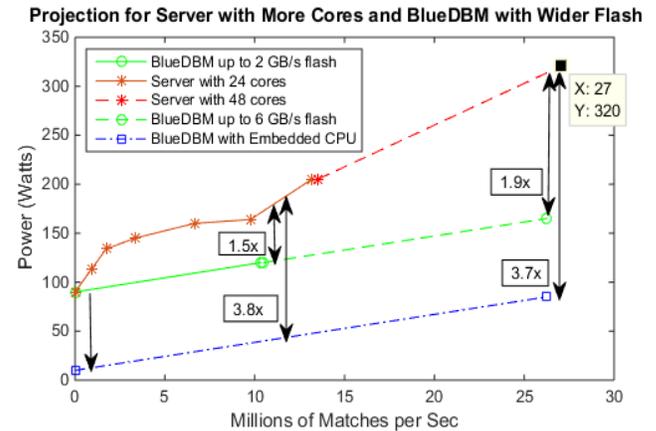

Fig. 14. Projection to Technology Generations

### C. Sparse Multiplications Performance

The formulation in Fig. 3 of Section II allows us to project the performance of our document matching kernel to a generalized performance metric of (nonzero) partial products per second. This metric is useful for estimating how applications such as topic modeling [17], centrality [18], and graph traversal that are amenable to sparse processing [5] could benefit from our accelerator.

In our experiment, processing of 8.2 million documents with 483 million words generates 11 million partial products and takes about 0.8 seconds. This yields **13 million partial products/sec** for sparse vector multiplication. The sparsity of the data is characterized by the number of prominent words out of a vocabulary set containing 141,000 words. A document, on average, contains about 60 such words, hence, a sparsity of 60/141,000 = 0.04%.

We could notionally attempt to compare our partial product processing rate to Graphulo server-side sparse matrix multiplication [19], keeping in mind 1) our results are immediately consumed – no need for saving back to database, and 2) the sparsity level should be similar to be relevant, although strictly speaking two datasets with similar sparsity may still have different number of partial products due to the actual distribution of nonzero terms. The data in the report had 16 nonzeros per vector. For vector size $2^{17} = 131,072$ ~141,000, the sparsity comes out to be 0.01%, not too far from our document search parameter.

Graphulo server-side multiplication peaks at about 300,000 partial products/sec per *pair* of CPU cores, attributed to architecture of the Tablet server. Assuming optimistic linear scaling with CPU cores, it would take a system with **86 CPU cores** to match the 13 million partial products/sec rate on one baseline BlueDBM accelerator slice, or **232 CPU cores** to match the optimized version.

## VI. SUMMARY

We have presented a prototyped document search accelerator using a novel in-storage processing architecture. We showed that the sparse pattern processing techniques can be extended to important applications including text analysis, bioinformatics, machine learning, graphs, etc. In our prototyping experiment, one baseline accelerator slice can process 10 million docs/sec, outperforming a 16-core system running C/C++ in-memory application, while incurring a fraction of the power and cost. An optimized accelerator could match a 48-core server system. Problem size up to 1 TB can be handled per accelerator slice.

There are many directions for future work. We would like to explore the application of in-storage analytics accelerator in a cloud [20][21] or supercomputing setting. We are also currently developing GraphBLAS compliant operations in our system for common graph and sparse linear algebra problems such as one proposed in [22]. We would also like to investigate how to integrate our system into more general data management solutions such as the BigDAWG polystore system [23]. Lastly, we are pursuing low SWaP computing solutions to enable selected subsets of the above to be processed in the field.